  \documentstyle[12pt]{article}
  
\catcode`@=11
\def\secteqno{\@addtoreset{equation}{section}%
\def\theequation{\thesection.\arabic{equation}}}
\catcode`@=12
\secteqno
\newcommand{\be}{\begin{equation}}
\newcommand{\ee}{\end{equation}}
\newcommand{\bea}{\begin{eqnarray}}
\newcommand{\eea}{\end{eqnarray}}
\newcommand{\bref}[1]{(\ref{#1})}
\newcommand{\nn}{\nonumber}

\newcommand{\slP}{/ {\hskip-0.27cm{P}}}

\newcommand{\slX}{/ {\hskip-0.27cm{X}}}
\newcommand{\slp}{/ {\hskip-0.27cm{p}}}
\newcommand{\sltp}{/ {\hskip-0.27cm{\tilde{p}}}}

\newcommand{\slSigma}{/ {\hskip-0.27cm{\Sigma}}}
\newcommand{\slL}{/ {\hskip-0.27cm{\bf L}}}
\newcommand{\slY}{/ {\hskip-0.27cm{Y}}}

\begin{document}
\thispagestyle{empty}
\vfill
\vbox{
\hfill October\ 23,\ 2000\null\par
\hfill KEK-TH-717\null\par
\hfill DAMTP-2000-115\null\par
}\null
\vskip 20mm
\begin{center}
{\Large\bf Open Superstring Theory and }\par
{\Large\bf Superalgebra of the Brane Antibrane System}\par
\vskip 10mm
{\large Machiko\ Hatsuda~~and~~Makoto\ Sakaguchi $^\dagger$}\par
\medskip
{\it 
Theory Division,\ High Energy Accelerator Research Organization (KEK),\\
\ Tsukuba,\ Ibaraki,\ 305-0801, Japan} \\
{\it 
$\dagger$ 
Department of Applied Mathematics and Theoretical  Physics\\
Wilberforce Road, Cambridge, CB3 0WA, U.K.}\\
\medskip
E-mail:\ mhatsuda@post.kek.jp,\ 
M.Sakaguchi@damtp.cam.ac.uk
\medskip
\vskip 10mm
\end{center}
\vskip 10mm
\begin{abstract}
Open superstring theory
is formulated in terms of a nondegenerate supertranslation algebra.
A supercharge for a tachyonic superstring 
can be also defined classically by taking into account the
leakage of the supercurrent which is compensated by fermionic and bosonic auxiliary fields.
The anticommutator of two 
supercharges
of the tachyonic superstring
does not contain the zero eigenvalue
and so this string is not a BPS state.
Brane-antibrane annihilation scenarios are described by these
superalgebras defined on the sum of
world-volumes of a D-brane, an anti-D-brane and a tachyonic superstring.
\end{abstract} 
\par
\newpage
\setcounter{page}{1}
\parskip=7pt
\section{ Introduction}\par
\indent

The classification of states according to their transformation by elements of
the superalgebra is a powerful tool for exploring non-perturbative aspects of supersymmetric theories \cite{WitOl}.
The anticommutator of two supercharges is a real symmetric matrix
\bea
\{Q_\alpha,Q_\beta\}=M_{\alpha\beta}~~~.
\eea
States are classified by the matrix $M_{\alpha\beta}$ as follows:
\begin{enumerate}
\item If all eigenvalues of $M_{\alpha\beta}$ are zero,
the state preserves all supersymmetries. 
This is the case for the ground states of fundamental superstrings,
i.e. the vacuum state which has no strings.
\item If at least one of eigenvalues of $M_{\alpha\beta}$ is zero,
the state preserves some supersymmetries.
This is a BPS state.
\item If all eigenvalues of $M_{\alpha\beta}$ are non-zero,
then all supersymmetries are broken.
This is a non-BPS state.
\end{enumerate}
As well as BPS systems \cite{Witb,pol},
recent developments of non-BPS systems have lead to the understanding
of a variety of non-perturbative phenomena in string theories
\cite{BaSus, GrGu,Sen4,BerG,Sen1,Sen2,Sen3,Horv,WitK}.

Although space-time supersymmetry (SUSY) is essential,
most studies of this field are developed by using 
the Neveu-Schwarz-Ramond (NSR) formalism rather than the 
Green-Schwarz (GS) formalism where
 space-time SUSY is manifest.
In order to represent space-time SUSY in the NSR formalism
a GSO projection is required,
and it plays an essential role in 
non-BPS physics.
The wrong GSO projection 
results in tachyonic modes
and
instability of the system \cite{Sen1}.
Recently it was pointed out that tachyonic modes are obtained 
in the GS
formalism by the ``wrong" boundary condition
by looking at the partition function \cite{Yone}.
In this paper
we try to explain appearance of tachyonic modes
in a GS like action.
Our guiding principle is the global SUSY of the action,
where SUSY is broken by the wrong boundary condition.
The problem is how to represent
the non-BPS superalgebra for a tachyonic superstring and how to recover SUSY.

In order to consider a tachyonic superstring
it is important to reexamine open string theory.
In the Kalb and Ramond discussion of the open string interaction \cite{KR}
the open string current is not conserved, and leaks from the end points.
The open string gauge interaction is not gauge invariant,
but gauge invariance is recovered by introducing a
vector gauge field
coupling with the endpoint currents.
For the type II superstring theories
the corresponding gauge invariant interaction is 
\bea
I_{WZ}=\frac{1}{2}\int_{\Sigma}B^{NS} -\int_{\partial \Sigma}A
~=~\frac{1}{2}\int_{\Sigma}{\cal B}~~~,~~~
{\cal B}=B^{NS}-dA~~~,\label{BandA}
\eea
where $\Sigma$ is
a string world-sheet, $\partial\Sigma$ is its boundary,
$B^{NS}$ is the Neveu-Schwarz two-form,
and $A$ is the Dirac-Born-Infeld (DBI) vector gauge field on Dp-branes
\cite{Witb}.
This Wess-Zumino action is not only gauge invariant but also SUSY invariant. 
The manifestly SUSY invariant Wess-Zumino action
can be obtained as an element of the trivial class of the Chevalley-Eilenberg cohomology of the nondegenerate supertranslation group \cite{Green,Siegel}.
The nondegenerate supertranslation group has been examined recently 
in many kinds of super-p-brane theories \cite{Sez,Sakag,Sakag1,AHKT,azc1,Sakag2,HS,HKS}.
The nondegenerate supertranslation algebra is obtained by introducing a fermionic central extension, and the fermionic center 
together with a supercharge makes the nondegenerate group metric.
This algebra may play a useful role in quantizing the theory
as shown in the random lattice case \cite{Siegel}.
In section 2 we construct an open superstring theory
based on the nondegenerate supertranslation algebra,
where fermionic and bosonic auxiliary fields carry 
boundary contributions.

For an open superstring there is an ambiguity in the GSO projection
in the NSR formalism.
In the GS formalism this ambiguity corresponds to 
an ambiguity in the boundary condition of fermionic coordinates \cite{Yone}.
A tachyonic state, which
has the wrong sign for the GSO projection, 
is obtained by considering a superstring connecting a Dp-brane and an anti-Dp-brane ($\overline{\rm Dp}$).
As an example if we take a superstring connecting a D2-brane and a $\overline{\rm D2}$-brane, then its boundary condition is given by
\footnote{
$\displaystyle\Gamma^{11}=\Gamma^\natural$,  
$\displaystyle\Gamma^\natural \pmatrix{\epsilon_1\cr \epsilon_2 }=\pmatrix{\epsilon_1\cr -\epsilon_2 }$
and $\Gamma_{012}=\Gamma_{[0}\Gamma_1\Gamma_{2]}/3!$.
}
\bea
\Gamma_{012}\theta_1&=&\theta_2~~~,~~~{\rm at}~ \sigma=0~~~\nn\\
\Gamma_{012}\theta_1&=&-\theta_2~~~,~~~{\rm at}~ \sigma=\pi~~~.
\label{thbd}
\eea
By defining
$\theta(\sigma)$ on $-\pi\leq\sigma\leq \pi$ 
by $\theta(\sigma)=\Gamma_{012}\theta_1(-\sigma)$ on 
$-\pi\leq \sigma\leq 0$ and
$\theta(\sigma)=\theta_2(\sigma)$ on $0\leq\sigma\leq \pi$,
$\theta$ is antiperiodic, $\theta(\pi)=-\theta(-\pi)$.
There is no global SUSY parameter satisfying \bref{thbd}
since $\theta(\sigma)$ has no zero mode.
This boundary condition still consistently cancels 
the surface term from the variation of
the light-cone gauge fixed action
\bea
\bar{\theta}_1\Gamma^-\delta \theta_1
-\bar{\theta}_2\Gamma^-\delta \theta_2=
\bar{\theta}_1\Gamma^-\delta \theta_1
-(\mp\bar{\theta}_1\Gamma_{210})\Gamma^-(\pm\Gamma_{012}\delta \theta_1)
=0
\eea
where $\Gamma^\pm=(\Gamma^0\pm \Gamma^1)/\sqrt{2}$
and $\Gamma^+\theta_{1,2}=0$.
\par
It turns out that 
the antiperiodic boundary condition on the fermionic coordinates
\bref{thbd}
leads to a non-zero value of the fermionic central charge
as we will see in
section 3.
We will also see that
the mass must have an imaginary value.
The non-BPS superalgebra,
which has no zero eigenvalue of $\{Q_\alpha,Q_\beta\}$,
is obtained by introducing an imaginary value for the tachyonic string 
energy
realizing complete SUSY breaking.
This choice corresponds to the wrong sign for the GSO projection.

In section 4, we demonstrate the D2-$\overline{{\rm D2}}$ annihilation
scenario from the superalgebra point of view. 
Depending on the configuration of the gauge fields
the D2- and $\overline{{\rm D2}}$-branes may annihilate
into nothing \cite{Sentch} or into D0-branes \cite{Senvo,WitK}.
After passing through unstable non-BPS states the system settles down 
in the stable BPS state.
The superalgebra gives a simple description
 of the brane charge cancellation and 
 the cancellation of the brane tensions and the tachyon potential.
\par

\section{Open superstring theory}
\indent

In this section we will obtain the gauge invariant
and SUSY invariant Wess-Zumino action for an open superstring
\bref{BandA}.
We first construct the nondegenerate supertranslation algebra
for type IIA as an example and then
calculate the SUSY invariant 
Maurer-Cartan (MC) one-forms
$L$,
and find a two-form ${\cal B}$ whose derivative gives 
the three-form $H=d{\cal B}$. 
Since the DBI gauge fields, with which the end points interact,
live in the branes
rather than in the string world-sheet,
the interaction is not written in terms of the string world-sheet
fields. 
Instead of using gauge fields in the Wess-Zumino action,
we enlarge the superspace and introduce auxiliary coordinates
which correspond to endpoint degrees of freedom.

The type IIA nondegenerate superalgebra \cite{Sakag}
for an open superstring
is 
\bea
&&\{Q_\alpha,Q_\beta\}=2
\left(C\slP +C\Gamma^\natural \slSigma\right)_{\alpha\beta}~~~\label{NDQQ}\\
&&
\left[ P_m,Q_\alpha \right]=i(Z\Gamma_m)_\alpha~~~\label{NDPQ}\\~~~
&&
\left[ \Sigma_m,Q_\alpha \right]=i(Z\Gamma^\natural\Gamma_m)_\alpha~~~,
\label{NDSQ}
\eea
which is determined from the 10-dimensional IIA cyclic identity
\bea
(\Gamma_m)^\alpha_{(\beta}(C\Gamma^m)_{\gamma\delta)}
+(\Gamma^{\natural}\Gamma_m)^\alpha_{(\beta}(C\Gamma^{\natural}\Gamma^m)_{\gamma\delta)}
=0~~~,~~~\Gamma^{\natural}=\Gamma^{11}.
\eea
The type IIB theory is obtained by replacing $\Gamma^\natural$
by $\tau_3$ acting on N=2 fermion index.
The enlarged superspace coordinates are introduced as parameters
of the group element
\bea
g=e^{i\xi^\alpha Z_\alpha}e^{iY^m \Sigma_m}
e^{iX^m P_m}e^{i\theta^\alpha Q_\alpha}
\eea
and the MC one-forms are obtained as
\bea
&&g^{-1}dg=i\left({\bf L}^mP_m+L^\alpha Q_\alpha +
{\bf L}_{\Sigma}^m\Sigma_m+L_Z^\alpha Z_\alpha\right)\nn\\
&&
\left\{\begin{array}{lcl}
{\bf L}^m&=&dX^m
-i\bar{\theta}\Gamma^m d\theta\\
 L^\alpha&=&d\theta^\alpha\\
{\bf L}^m_\Sigma&=&dY^m
-i\bar{\theta}\Gamma^\natural\Gamma^m d\theta\\
 L^\alpha_Z&=&d\xi^\alpha
+ (\Gamma_m\theta)^\alpha
 (dX^m
 -i\displaystyle\frac{1}{3}\bar{\theta}\Gamma^md\theta)\\
 &&~~~~
 +( \Gamma^\natural\Gamma_m\theta)^\alpha
 (dY^m
 -i\displaystyle\frac{1}{3}\bar{\theta}
 \Gamma^\natural\Gamma^md\theta)
\end{array}\right.\label{MC1}
\eea
which satisfies the MC equations
\bea
&&d{\bf L}^m
+i\bar{L}\Gamma^m L=0~~~,~~~~dL=0\nn\\
&&d{\bf L}^m_\Sigma
+i\bar{L}\Gamma^\natural\Gamma^m L=0~~,~~~
dL_Z
-\slL L
-\Gamma^\natural \slL_\Sigma L=0~~~.
\eea
The NS two-form ${\cal B}$ whose derivative becomes closed three-form $H$
is given by 
\bea
d{\cal B}&=&H~=~\bar{L}\Gamma^\natural \slL L\nn\\
{\cal B}&=&\frac{
1}{2}(\bar{L}\Gamma^\natural L_Z
-i{\bf L}\cdot{\bf L}_\Sigma)\label{BmdA}~~~,
\eea
and this is the combination which enters into
the SUSY invariant Wess-Zumino action.

An action for the open superstring theory of type IIA 
is
\bea
I_{\rm F1'}&=&
\int d^2\sigma({\cal L}_{NG}+{\cal L}_{WZ})\label{IF1I}\\
{\cal L}_{NG}&=&-T\sqrt{-\det G_{\mu\nu}}~~
,~~G_{\mu\nu}={\bf L}_\mu\cdot {\bf L}_\nu\nn\\
{\cal L}_{WZ}&=& -\displaystyle\frac{i}{2}
T\epsilon^{\mu\nu}{\cal B}_{\mu\nu}~~,~~
{\cal B}_{\mu\nu}=%
-\frac{1}{2}
(\bar{L}_{\mu}\Gamma^\natural
L_{Z\nu }
+i{\bf L}_{\mu}\cdot{\bf L}_{\Sigma \nu})-(\mu\leftrightarrow\nu)
\label{IF1L}
\eea 
where $\sigma^\mu$ are worldsheet coordinates with $\mu=0,1$.
The two form,
${\cal B}=\displaystyle\frac{1}{2}d\sigma^\mu
d\sigma^\nu {\cal B}_{\mu\nu}$,
 is given by \bref{BmdA} and
\bea
&&~~{\cal B}_{\mu\nu}=B_{\mu\nu}-\partial_{[\mu}A_{\nu ]}\nn\\
&&\left\{\begin{array}{ccl}
B_{\mu\nu}&=&\partial_\mu\bar{\theta}\Gamma^\natural \Gamma^m\theta
(\partial_\nu X_m
-i\displaystyle\frac{1}{2}\partial_\nu\bar{\theta}\Gamma_m\theta)-(\mu\leftrightarrow\nu)
\\
A_\nu&=&\displaystyle\frac{1}{2}(
\bar{\theta}\Gamma^\natural\partial_\nu\xi
+iX^m\partial_\nu Y_m)
\end{array}\right.\label{WZBA}~~~.
\eea
The constraint set of the system is obtained as
\bea
H&=&\tilde{p}^2+T^2G_{11}=0\label{bconH}\\
H_\perp&=&\tilde{p}\cdot {\bf L}_1=0\label{bconHp}\\
F&=&\zeta
-i\bar{\theta}(\sltp-\displaystyle\frac{T}{2}\slL_{\Sigma 1}
+\displaystyle\frac{T}{2}\Gamma^\natural\slL_1)
+i\displaystyle\frac{T}{2}\bar{L}_{Z1}\Gamma^\natural\nn\\
&&
+\displaystyle\frac{T}{6}(\bar{L}_1\Gamma^\natural\Gamma\theta
\cdot\bar{\theta}\Gamma+\bar{L}_1\Gamma\theta\cdot\bar{\theta}\Gamma^\natural\Gamma)=0
\label{fercon}\\
\phi_Y&=&p_Y-\displaystyle\frac{T}{2}
({\bf L}_1
+i\bar{L}_1\Gamma\theta)=
p_Y-\displaystyle\frac{T}{2}X'
=0
\label{bconp}\\
\phi_\xi&=&\pi_\xi+i\displaystyle\frac{T}{2}
\bar{L}_1\Gamma^\natural=\pi_\xi+i\displaystyle\frac{T}{2}
\bar{\theta}'\Gamma^\natural=0
\label{fconp}
\eea
for canonical pairs $(X^m,\theta^\alpha,Y^m,\xi^\alpha)$ and
$(p_m,\zeta_\alpha,p_{Ym},\pi_{\xi \alpha})$ and where
$\tilde{p}$ is the SUSY invariant combination
\bea
\tilde{p}_m=p_m+\displaystyle\frac{T}{2}{\bf L}_{\Sigma 1 m}
-i\displaystyle\frac{T}{2}\bar{L}_1
\Gamma^\natural\Gamma_m\theta=
p_m
+\displaystyle\frac{T}{2}Y'_m
+iT\bar{\theta}\Gamma^\natural\Gamma_m\theta'
~~~.
\eea
The constraint set is invariant under the global SUSY transformation 
\bref{SUSYrule} up to  constraints.
As in \cite{HKS}
 the fermionic constraints \bref{fercon} satisfies the
following algebra 
\bea
\{F_\alpha(\sigma),F_\beta(\sigma')\}=-2i(C\Xi_+)_{\alpha\beta}
\delta(\sigma-\sigma')~~~,~~~
\Xi_\pm=\Gamma^m\tilde{p}_m\pm T\Gamma^\natural\Gamma^m{\bf L}_{1m}
~~~.\label{FFkappa}
\eea
One half of $F_\alpha$ are first class constraints
generating the kappa symmetry with generator
given by
\bea
\tilde{F}_\alpha=({F\Xi_-})_\alpha=0~~~,~~~\Xi_+\Xi_-\approx 0~~~.
\label{Gkappa}
\eea

It is important to examine behavior of the endpoints.
For simplicity,
let us consider the ground state in which fermionic variables can be neglected.
For the Neumann boundary directions denoted by $\mu$,
${\bf L}_{1}^\mu=0$ is satisfied at boundaries.
For the Dirichlet boundary directions denoted by $i$, 
${\bf L}_{1}^i\neq 0$ at boundaries is obtained.
From the constraint \bref{bconHp}
it follows that $\tilde{p}^i=0$.
The $\tau$ diffeomorphism constraint \bref{bconH} 
at the boundaries is written as
\bea
H
=({\tilde{p}}^\mu)^2+T^2(X'^i)^2=0~~~.
\eea
The Dirichlet boundary gives rise to a mass $T$.
As a result massive endpoints are moving on D2-branes.

The SUSY transformations are given by
\bea
\left\{\begin{array}{lcl}
\delta \theta^\alpha&=&\epsilon^\alpha\\
 \delta X^m&=&-i\bar{\epsilon}\Gamma^m\theta\\
 \delta Y^m&=&-i\bar{\epsilon}\Gamma^\natural \Gamma^m\theta\\
 \delta \xi^\alpha&=&\slX\epsilon^\alpha +\Gamma^\natural\slY \epsilon^\alpha
 -\displaystyle\frac{i}{3}(\Gamma\theta^\alpha\cdot \bar{\epsilon}\Gamma \theta
 +\Gamma^\natural\Gamma\theta^\alpha\cdot \bar{\epsilon}\Gamma^\natural
 \Gamma \theta)
\end{array}\right.~~~,\label{SUSYrule}
\eea
under which the MC one-forms \bref{MC1} are invariant.
The general form of the supercharges is
\bea
Q\epsilon&=&\int d\sigma \left(
p\cdot\delta X+\zeta \delta\theta+p_Y\cdot\delta Y+
\pi_\xi \delta\xi
\right)\label{Qe}~~~.
\eea
Using \bref{SUSYrule}
the supercharges are obtained as
\bea
Q_\alpha&=&
\int d\sigma \left[
\zeta 
+i\bar{\theta}\slp
+i\bar{\theta}\Gamma^\natural\slp_Y
+\pi_\xi(\slX+\Gamma^\natural\slY)
+i\displaystyle\frac{1}{3}
(\pi_\xi \Gamma\theta\cdot \bar{\theta}\Gamma
+\pi_\xi \Gamma^\natural\Gamma\theta\cdot \bar{\theta}
\Gamma^\natural\Gamma)
%
\right]_\alpha~~~\label{Super}
\eea
which satisfy the expected superalgebra \bref{NDQQ}
\bea
\{Q_\alpha,Q_\beta\}=2i\left(C\slP 
+C\Gamma^\natural \slSigma
\right)_{\alpha\beta}~~~\label{QQpd}.
\eea
The factor of $i$ on
the right hand side 
arises by the usual replacement of
the commutator,
$[X^m(\sigma),P_n(\sigma')]=-i\delta^m_n\delta(\sigma-\sigma')$,
by
the Poisson bracket,
$\{X^m(\sigma),P_n(\sigma')\}=\delta^m_n\delta(\sigma-\sigma')$.

Other charges are written in terms of canonical variables
\bea
P_m&=&\int d\sigma~ p_m\\
\Sigma_m&=&\int d\sigma ~p_{Ym}\approx
\displaystyle\frac{T}{2}\int d\sigma~ X'_m
=-\displaystyle\frac{T}{2}(X_m|_{\sigma=0}-X_m|_{\sigma=\pi})
\label{sigmax}
\\
Z_\alpha&=&\int d\sigma ~\pi_{\xi \alpha}
\approx
-i\displaystyle\frac{T}{2}\int d\sigma ~\bar{\theta}'\Gamma^\natural_\alpha
=i\displaystyle\frac{T}{2}( \bar{\theta}\Gamma^\natural_\alpha
|_{\sigma=0}- \bar{\theta}\Gamma^\natural_\alpha
|_{\sigma=\pi})
\eea
where constraints \bref{bconp} and \bref{fconp} are used.
The periodic boundary condition gives
\bea
Z_{\alpha}=0~~~.\label{Zzero}
\eea
This is consistent with the algebra \bref{NDPQ} and 
\bref{NDSQ} in the presence of
D-branes where $P_i=0,~\Sigma_i=0$, $i=4,...,9$ in 
the Dirichlet directions.


If we use the constraint \bref{bconp}, the BPS condition obtained
from the superalgebra \bref{NDQQ} is different from the expected one;
the BPS mass is half of the expected value, $P_0=Td/2$.
The origin of factor $1/2$ is the coefficient of the Wess-Zumino action
which is determined by the kappa symmetry \bref{FFkappa} and \bref{Gkappa}.
To avoid confusion,
note that this mass, $P_0=Td/2$,
is the mass of an open superstring attached to two branes
including the
effect of the auxiliary fields.
If we impose $P_0=Td$
the coefficient of the Wess-Zumino action should be a factor of 2 larger.
This breaks the kappa symmetry and there are twice as many fermionic
variables.
This phenomenon is a property of the world-volume action
 for non-BPS branes
in \cite{Senwv}. 
In this paper we will consider a situation
in which the BPS condition is modified in a manner that preserves the kappa symmetry.
This will allow an efficient
first-quantized description of tachyon condensation.

\par
\section{The tachyonic superstring 
}
\indent

In this section we focus on an open superstring connecting 
D2- and $\overline{\rm D2}$-branes with separation $d$.
The fermionic coordinates $\theta$ satisfy \bref{thbd}
which is an antiperiodic boundary condition.
It is interesting to consider this situation
from the superalgebra point of view. 
We begin with the nondegenerate supertranslation algebra
for an open superstring \bref{NDQQ}, \bref{NDPQ} and \bref{NDSQ}.

When the antiperiodic boundary condition \bref{thbd} is imposed,
the fermionic central charges cannot be set to be zero;
\bea
Z_{A\alpha}&\equiv&-i\displaystyle\frac{T}{2}( \bar{\theta}_A\Gamma^\natural_\alpha
|_{\sigma=0}- \bar{\theta}_A\Gamma^\natural_\alpha
|_{\sigma=\pi})\nn\\
Z_{1\alpha}&=&0,~\to ~Z_{2\alpha}~\neq~0
 ~~~.\label{Znonzero}
\eea
The nonvanishing fermionic center $Z_\alpha$ in \bref{Znonzero}
is crucial in this non-BPS system in contrast with \bref{Zzero}
which describes a BPS system.
This nonvanishing property leads to an inconsistency in the algebra \bref{NDPQ} and \bref{NDSQ} or equivalently
\bea
\left[P_m-\Sigma_m,Q_1\right]&=&2iZ_1\Gamma_m~=~0\nn\\
\left[P_m+\Sigma_m,Q_2\right]&=&2iZ_2\Gamma_m~\neq~0\label{PSQZ}~~~.
\eea
Although the right hand sides of \bref{PSQZ} are non-zero,
the Dirichlet components of
the left hand side are zero 
for $d\to 0$ because 
$P_i=0$ and 
$\Sigma_i=0$, 
 $i=3,...,9$.
In other words, by taking the limit $d\to 0$, the antiperiodic boundary condition on $\theta$'s, twisting fermions, causes a
singularity in the superspace. 
One resolution of this inconsistency is 
to consider complex $P_m$ 
,
which allows for the presence of a tachyonic open string state.
It is expected that this singularity is resolved by quantum excitation
of such tachyonic modes.
Collective tachyonic excitation makes the tachyon potential
which contributes to the energy, $P_0$.
This will be discussed in the next section.
In this section we concentrate on the algebraic consistency,
where we introduce ``tachyonic modes" in the algebra
as ``imaginary" mass,
$P=\Re P+i\Im P\neq 0$.
So the left hand side of \bref{PSQZ} becomes non-zero,
$(P+\Sigma)=\Re (P+\Sigma)+i\Im P\neq 0$.
The BPS condition should now be examined carefully,
by considering the superalgebra \bref{NDQQ}
for a string lying along $X^3$ direction at rest
\bea
\{Q,Q\}\epsilon&=&2(P_0-\displaystyle\frac{Td}{2}\Gamma^{\natural 03})\epsilon=0~~~.\label{QQ03}
\eea
With imaginary $P_m$, say 
$P_0=im_T$ with real 
 $m_T$,
\bea
\det (
im_T-\frac{Td}{2}\Gamma^{\natural 03})
\neq 0~~~,
\eea
so the only solution of (3.3) is $\epsilon=0$ and SUSY is completely broken.
In this way the antiperiodic boundary condition leads to a
non-BPS algebra.
Note that in this classical approach, 
the value of the first quantized tachyonic mass
$m_{Tachyon}^2=-\frac{1}{2\alpha'}$ 
cannot be determined.

It is important that
there is a supercharge \bref{Super} 
under which the action is invariant not only for the
usual fundamental open superstring but also a tachyonic superstring.
The use of auxiliary fields is essential here to make a self-consistent 
open string theory.
Without introducing local auxiliary fields,
the global symmetries of the open string action depend on
the boundary interactions.
By using auxiliary fields the SUSY invariance 
 can be guaranteed at least at the level of the action,
 independent of the boundaries.
The difference between the usual superstring and
the tachyonic superstring
is  ``the expectation value" of the fermionic central charge $Z_\alpha$
and momentum $P_m$.
Once the vacuum is chosen, ``the expectation value"
is fixed and the extended superalgebra
\bref{NDQQ}, \bref{NDPQ} and \bref{NDSQ} 
leads to BPS states or non-BPS states as shown here.    

\par

\section{The Dp-$\overline{\rm{\bf Dp}}$ system}
\indent

Now let us consider the Dp-F1-${\overline{\rm Dp}}$ system. 
As an example we consider the following configuration of type IIA theory :
A D2-brane and a ${\overline{\rm D2}}$-brane lying in the $X^1$-$X^2$ directions,
and an open superstring with Dirichlet boundaries  
$X^i=0$ with $i=3,...,9$ at $\sigma=0$ 
and $X^3=d$ and $X^i=0$ with $i=4,...,9$ at $\sigma=\pi$.
The total action of this system is given by the sum of actions for 
a D2-brane, a $\overline{\rm D2}$-brane and an
open superstring (F1) and their interactions:  
\bea
I_{total}&=&I_{{\rm D2}}[A_\mu]+I_{\overline{\rm D2}}[\bar{A}_\mu]+I_{\rm F1}+I_{int}[A_\mu,{\bar{A}}_\mu]\nn\\
I_{int}&=&\int_{{\rm D2}} d\tau A_\mu j^\mu|_{\sigma=0}-
\int_{\overline{\rm D2}} d\tau {\bar{A}}_\mu j^\mu|_{\sigma=\pi}~~~.
\label{Itotal}
\eea 
The end points of an open superstring are sources of DBI gauge fields,
so the Gauss law constraints become
\bea
\partial_aE^a=j^0|_{\sigma=0}=T\delta^{(3)}(x-X(\tau))~,~
\partial_a{\bar{E}}^a=-j^0|_{\sigma=\pi}=-T\delta^{(3)}(x-X(\tau))~,
~a=1,2
\eea
where the canonical conjugates of $A_\mu$, ${\bar{A}_\mu}$ are 
$E^\mu$, ${\bar{E}}^\mu$, $\mu=0,1,2$.
Under a global SUSY transformation, 
$I_{{\rm D2}}$, $I_{\overline{\rm D2}}$ and $I_{\rm F1}+I_{int}$ are invariant.
Although the Wess-Zumino action of an open superstring $I_{\rm F1}$ 
gives rise to a nonvanishing surface term,   
the variation of the DBI gauge fields in the
interaction $I_{int}$ cancels it.
We identify this SUSY invariant combination $I_{\rm F1}+I_{int}$
with the self-contained open superstring action $I_{\rm F1'}$ 
\bref{IF1L}.
Auxiliary fields in $I_{\rm F1'}$ are related to the DBI gauge fields
by the relation \bref{WZBA}.
It may be confusing that there exist Wess-Zumino actions not only for
an open superstring but also Dp,$~\overline{\rm Dp}$ branes,
but in this paper we modify 
only a Wess-Zumino action for an open superstring
and keep usual forms of Wess-Zumino actions 
for Dp,$~\overline{\rm Dp}$ branes 
which have usual boundary conditions.

Each Noether charge on world-volume
is written in terms of
world-volume variables,
so different world-volume charges commute/anticommute
as long as the world-volumes are separated.
The superalgebra of a D2-brane is given by \cite{MK}, 
\bea
\{Q_{{\rm D2}\alpha},Q_{{\rm D2}\beta}\}&=&2i\left(C\slP-T_{D2} C\int \partial_1\slX \partial_2\slX\right)_{\alpha\beta}\nn
+2i\left( C\Gamma^\natural \int \partial_a \slX E^a
-C\Gamma^\natural T_{D2}\int  F_{12} \right)_{\alpha\beta}\\
&&+\frac{i}{2}\int  \partial_aE^a
\left( 
\bar{\theta}\Gamma^\natural \Gamma_\alpha\cdot\bar{\theta}\Gamma_\beta
+\bar{\theta}\Gamma^\natural\Gamma_\beta\cdot\bar{\theta} \Gamma_\alpha\right)
\label{QD2}
\eea
An anti-BPS state is obtained by reversing the brane charge,
or equivalently by taking opposite sign for the world-volume coordinate, 
for example
\bea
&&\displaystyle\int_{\overline{\rm D2}} 
d^2\sigma \epsilon^{ab} T_{\rm D2}\partial_a \slX \partial_b \slX
= -\displaystyle\int
d^2\sigma \epsilon^{ab} T_{\rm D2}\partial_a \slX \partial_b \slX
=-T_{\rm D2}V\nn\\
&&\displaystyle\int_{\overline{\rm D2}} 
d^2\sigma \epsilon^{ab} F_{ab}
= -\displaystyle\int
d^2\sigma \epsilon^{ab} F_{ab}~~~.
\eea
The superalgebra for an anti-D2-brane is given by
\bea
\{Q_{\overline{\rm D2}\alpha},Q_{\overline{\rm D2}\beta}\}
&=&
2i\left(C\slP-T_{D2} C\int_{\overline{\rm D2}}  \partial_1\slX \partial_2\slX\right)_{\alpha\beta}\nn
+2i\left( C\Gamma^\natural \int_{\overline{\rm D2}} \partial_a \slX {\bar{E}}^a
-C\Gamma^\natural T_{D2}\int_{\overline{\rm D2}}  {\bar{F}}_{12} \right)_{\alpha\beta}\\
&&+\frac{i}{2}\int_{\overline{\rm D2}}  \partial_a{\bar{E}}^a
\left( 
\bar{\theta}\Gamma^\natural \Gamma_\alpha\cdot\bar{\theta}\Gamma_\beta
+\bar{\theta}\Gamma^\natural \Gamma_\beta\cdot\bar{\theta}\Gamma_\alpha
\right)\nn\\
&=&
2i\left(C\slP+T_{D2} C\int
  \partial_1\slX \partial_2\slX\right)_{\alpha\beta}\nn
+2i\left( C\Gamma^\natural \int \partial_a \slX {\overline{E}}^a
+C\Gamma^\natural T_{D2}\int  {\bar{F}}_{12} \right)_{\alpha\beta}\\
&&+\frac{i}{2}\int \partial_a{\overline{E}}^a
\left( 
\bar{\theta}\Gamma^\natural \Gamma_\alpha\cdot\bar{\theta}\Gamma_\beta
+\bar{\theta}\Gamma^\natural \Gamma_\beta\cdot\bar{\theta}\Gamma_\alpha
\right)
\label{QaD2}~~~.
\eea
 The expression for the superalgebras for other Dp-brane systems are given in
\cite{HK,KH,MK}.

We focus on one superstring connecting unit volumes of the D2-brane
and $\overline{\rm D2}$-brane.
The total supercharge is 
\bea
Q&=&Q_{{\rm D2}}+Q_{\overline{\rm {\rm D2}}}+Q_{\rm F1'}~~~.\label{totalQ}
\eea
First we look for a case where the DBI gauge fields 
do not have any topological excitation.
The anticommutator of the total supercharges \bref{totalQ} is
\bea
\{Q,Q\}&=& \{Q_{{\rm D2}},Q_{{\rm D2}}\}
+\{Q_{\overline{\rm D2}},Q_{\overline{\rm D2}}\}
+\{Q_{\rm F1'},Q_{\rm F1'}\}\nn\\
&=&
2\left((P_{{\rm D2},0}+P_{\overline{\rm D2},0}+P_{F1',0})
-(T_{{\rm D2}}-T_{{\rm D2}} ) \Gamma^{012}
+\displaystyle\frac{Td}{2} \Gamma^{\natural 03}
\right)
\label{QQtotal}
~~~.
\eea
The D2- and ${\overline{\rm D2}}$-brane charges in the second term cancel.
The $F1'$ charge in the third term vanishes in the limit $d\to 0$.
When we bring the $\overline{\rm D2}$-brane close to 
the D2-brane ($d\to 0$),
many tachyonic modes are excited and form the tachyon potential.
We treat one tachyonic superstring in the background of
many other tachyonic superstrings as
a string in a tachyon potential, 
using the mean field approximation.
Whereas in section 3 we considered a single tachyonic superstring
we will now discuss a tachyonic string in the presence of a tachyonic potential
in order to describe tachyon condensation.
The energy for an isolated tachyonic string,
$P_{\rm F1',0}=im_T$,
corresponds to the top of the 
local maximum of the tachyon potential, 
 while the energy
 in the presence of
 $N$ tachyonic strings is 
 $\displaystyle\sum^N P_{\rm F1',0}=N{\cal V}_{min}({\cal T})$
 which corresponds to the local minimum of the tachyon potential.
The average energy of such a tachyonic superstring is 
\bea
P_{\rm F1',0}
{\longrightarrow}
{\cal V}_{min}({\cal T})~~~.
\eea
The energy per unit volume of the D2-($\overline{\rm D2}$-)brane
is $P_{D2,0}=T_{D2}$ ($P_{\overline{\rm D2},0}=T_{D2}$).
The condition of the first term in \bref{QQtotal} 
\bea
P_{{\rm D2},0}+P_{\overline{\rm D2},0}+P_{\rm F1',0}=
2T_{\rm D2}+{\cal V}_{min}({\cal T})=0
\eea
describes the situation in which a 
D2-${\overline{\rm D2}}$ pair annihilates.

Next the gauge field contribution to 
the superalgebra \bref{QQtotal} is taken into account;
\bea
\{Q_\alpha,Q_\beta\}&=& 
2\left((P_{{\rm D2},0}+P_{\overline{\rm D2},0}+P_{F1',0})
-(T_{{\rm D2}}-T_{{\rm D2}} ) \Gamma^{012}
+\displaystyle\frac{Td}{2} \Gamma^{\natural 03}
\right)\nn\\
&&
+2\left( \Gamma^{0\natural a}  \int( E^a+ \bar{E}^a)
-\Gamma^{0\natural}  T_{{\rm D2}}\int( F_{12}-\bar{F}_{12}) \right)\nn
\\
&&+\int \partial_a (E^a + \bar{E}^a)
\frac{i}{2}\left( 
\bar{\theta}\Gamma^\natural \Gamma_\alpha\cdot\bar{\theta}\Gamma_\beta
+\bar{\theta}\Gamma^\natural\Gamma_\beta\cdot\bar{\theta} \Gamma_\alpha\right)
\label{QQtotalEF}~~~.\nn\\&&
\eea
After computing the algebra, 
the integral on the right hand side of \bref{QQtotalEF}
can be written as a 2-dimensional world-volume integral.
When the vortex configuration occurs in the
tachyon fields, ${\cal T}=|{\cal T}|e^{in\chi}
$,
the magnetic fields on the D2-branes 
take the following configurations  
\bea
F_{12}-\bar{F}_{12}=\frac{1}{2\pi}[\partial_1, \partial_2] \ln {\cal T}~~~.\label{FFvor}
\eea
This tachyon vortex configuration does not give
an electric configuration,
so we may set
$E^a + \bar{E}^a=0$. 
However, it is also interesting to consider nontrivial configurations of 
$E+\bar{E}$ \cite{Yi}.
By taking the 2+1 space-time dual, 
$\epsilon_{ab}(E^a + \bar{E}^a)=\partial_b(\varphi-\bar{\varphi})$,
scalar fields $\varphi$ and $\bar{\varphi}$ appear
which correspond to 11-th dimensional coordinates.
So nonzero $\int(E^a + \bar{E}^a)$ means  
nontrivial winding of the relative D2-$\overline{\rm D2}$ 
(M2-$\overline{\rm M2}$) on the compact 11-th dimensional direction.
Since the term
$\Gamma^{0\natural  a}\displaystyle\int(E^a + \bar{E}^a)$ 
in \bref{QQtotalEF} 
is universal for arbitrary Dp-branes
\cite{HK,KH,MK},
the superalgebras for an arbitrary Dp-$\overline{\rm Dp}$ system
are sensitive to this type of winding effect.

As a result of using \bref{FFvor} the superalgebra is
\bea
\{Q,Q\}&=& 
2(P_{0}-nT_{{\rm D2}}\Gamma^{0\natural})
\label{QQtotalvor}~~~,
\eea
which is nothing but the superalgebra for $n$ D0-branes.
Half the SUSY is recovered, and the final state is a BPS state.


\section{Discussion}\par
\indent

We have shown that the nondegenerate supertranslation algebra
is suitable for representing 
an open superstring including the tachyonic superstring.
The supercharge
can be defined,
taking into account the
leakage of the supercurrent which is compensated by fermionic and bosonic auxiliary fields.
For a tachyonic superstring
the antiperiodic boundary condition on fermionic coordinates 
leads to a nonzero fermionic central charge and it turns out that imaginary valued momentum 
is required.
In other words, both the fundamental superstring and the tachyonic 
superstring have the same symmetry structure and the difference is 
in the 
expectation value of the global charges.   
The GSO ambiguity in the NSR formalism  
corresponds to the ambiguity of the expectation values
of momentum $P$
in the GS formalism
as is seen for the BPS condition \bref{QQ03},
\bea
\{Q,Q\}\epsilon=2(P_0-\frac{Td}{2}\Gamma^{\natural 03})\epsilon=0~~~.
\eea
This allows the following possibilities.
\bea
\begin{array}{c|c|c|c|c|c}
 &{\rm Mass}&{\rm SUSY}&\{Q_\alpha,Q_\beta\}& \theta~{\rm boundary}&{\rm GSO~}\\
 \hline
{\rm BPS}& P_0=\displaystyle\frac{Td}{2} &
 \begin{array}{c}
 \epsilon=\Gamma^{\natural 03}\epsilon\\\to ^\exists \epsilon~
 \end{array}
 &
  \pmatrix{{\bf 0}&{\bf 0}\\ {\bf 0}&{\bf 2}} & 
{\rm periodic}& {\rm correct}\\
\hline
  {\rm non\mbox{-}BPS}&P_0=i
  m_T& 
  \begin{array}{c}
  \epsilon=0
  \end{array}
  & 
\det\{Q_\alpha,Q_\beta\}\neq 0
&
{\rm antiperiodic}& {\rm wrong}
\end{array}\nn
\eea

We also showed that the superalgebra, defined by the sum of world-volumes of D2-F1'-$\overline{\rm D2}$, represents brane-antibrane annihilation.
Without a topologically nontrivial configuration of tachyon fields, 
the brane-antibrane system which originally contains
the non-BPS tachyonic superstring, reduces to nothing, which is 
the supersymmetric vacuum.
If the tachyonic fields are in a vortex configuration,
the brane-antibrane system reduces to 
the D0-brane system which is a BPS state.
The merit of the superalgebra is that
important combinations of charges of the system 
are automatically derived;
such as the cancellation of the D-brane tensions,
electromagnetic quantum numbers, 
$\int(F_{ab}-\bar{F}_{ab})$ and $\int(E^a+\bar{E}^a)$.
A problem of this approach is that
the quantum tachyon mass or
the field theoretical tachyon potential
cannot be deteremined.
This approach to the brane-antibrane system 
from the point of view of space-time SUSY
is purely classical.

Applications of the non-degenerate superalgebra to manifestly supersymmetric 
open membrane theory and other p-brane theories 
are also possible
\cite{Sez}.
The extension to non-abelian gauge interactions is an important issue.
In this formulation the vector gauge interaction 
is represented by using auxiliary fields,
so 
it is expected that the Chan-Paton factor is simply imposed on
the auxiliary fields.
This approach, where the space-time symmetry is manifest,
will be useful for examining the conjecture that
all branes can be obtained from the D9- and ${\overline{\rm D9}}$-brane
\cite{WitK}.
We leave these problems for a further publication.

\vskip 6mm
{\bf Acknowledgments}\par
We would like to express our gratitude to Michael B. Green
for careful reading the manuscript and useful discussions.  
M.H. would like to thank Tamiaki Yoneya for a seminar given at KEK
and for fruitful discussion,
and she wishes to thank
 Ken-ji Hamada, Masashi Hayakawa and Nobuyuki Ishibashi for
useful discussions.
M.S. wishes to thank YITP and the theory group of KEK for 
the kind hospitality
and gratefully acknowledges support
by Nishina Memorial Foundation.

\end{document}